\documentstyle[prb,aps,twocolumn,psfig,floats]{revtex}

  \def\esp#1{e^{\,\textstyle#1}}

  \def\relaz#1#2{\mathrel{\mathop{\kern 0pt#1}\limits_{#2}}}

\begin{document}
\wideabs{
\title{Finite-temperature ordering in two-dimensional magnets}
\author{Alessandro Cuccoli, Valerio Tognetti, Tommaso Roscilde and Paola
Verrucchi}
\address{Dipartimento di Fisica dell'Universit\`a di Firenze
        and Istituto Nazionale di Fisica della Materia (INFM),
        \\ Largo E. Fermi~2, I-50125 Firenze, Italy}

\author{Ruggero Vaia}
\address{Istituto di Elettronica Quantistica
        del Consiglio Nazionale delle Ricerche,
        via Panciatichi~56/30, I-50127 Firenze, Italy}
\date{\today}

\maketitle

\begin{abstract}

We study the two dimensional quantum Heisenberg antiferromagnet on the square
lattice with easy-axis exchange anisotropy.
By the semiclassical method called
pure-quantum self-consistent harmonic approximation we analyse several
thermodynamic quantities and investigate the existence of a finite
temperature transition, possibly describing the low-temperature critical
behaviour experimentally observed in many layered real compounds. 

We find that an Ising-like transition characterizes the model even when
the anisotropy is of the order of $10^{-2}J$ ($J$ being the intra-layer
exchange integral), as in most experimental situations. 
On the other hand, typical features of the isotropic Heisenberg 
model are observed for both values of anisotropy considered, 
one in the {\it quasi}-isotropic limit and the other in a more markedly 
easy-axis region.

The good agreement found between our theoretical results 
and the experimental data relative to the real compound Rb$_2$MnF$_4$
shows that the insertion of the easy-axis exchange anisotropy,
with quantum effects properly taken into account, 
provides a quantitative description and explanation of the experimental data,
thus allowing to recognize in such anisotropy the 
main agent for the observed onset of finite temperature long-range 
order.

\end{abstract}

}

\section{Introduction}
\label{s.intro}

In recent years the increasing interest in the physics of low-dimensional
magnets has led to a deep analysis of the thermodynamic behaviour of
two-dimensional quantum antiferromagnets\cite{deJongh90} and, in particular, of the 
isotropic quantum Heisenberg antiferromagnet on the square lattice
(QHAF)\cite{Manousakis91,KastnerEtal98}. Such model has in fact been widely used to describe the magnetic
properties of many quasi two-dimensional
real compounds, from the $S=1/2$ cuprate La$_2$CuO$_4$
\cite{ShiraneEtal87,KeimerEtal92} to the $S=5/2$ compound
Rb$_2$MnF$_4$, recently studied by Lee et al.\cite{LeeEtal98}.

Different theoretical
methods \cite{ArovasA88,Takahashi89,ChakravartyHN89,HasenfratzN91,ElstnerEtal95,CTVV9697iso} 
have been used to examine the rich reservoir of experimental data and the picture
of the subject is now well focused, albeit with some shaded parts.
In particular open questions still
exist\cite{BeardEtal97,KimT98,Hasenfratz00,BeardCK00}
on the low-temperature region, where the spin correlation length becomes 
of the order
of $10^2$ lattice spacings and the real magnets are seen to develop 
macroscopic areas of correlated spins. 

The theoretical debate on the low-temperature regime has been mainly 
dedicated to the isotropic QHAF, but 
real compounds are not actually well described by such isotropic model 
when temperature is lowered: the experimental
evidence of a finite-temperature transition, opposite to the Mermin-Wagner
theorem\cite{MerminW66} assertion that such a transition cannot occur in 
the two-dimensional 
isotropic QHAF, suggests that three-dimensional correlations and possible
anisotropy effects, as well as a combination of both,  must be considered.

The magnetic structure of layered real compounds is 
such that the exchange integral $J$ for neighbouring spins belonging to 
the same plane is orders of magnitude larger than that 
for neighbouring spins on different planes
\cite{Lines67,BirgeneauGS69,BirgeneauGS70}, hereafter called $J'$; 
one would hence naively expect the magnetic properties to be 
those of an effective two-dimensional magnet down to 
temperatures of the order of $J'$, until the 
transition towards an ordered three-dimensional phase should take place.
However, the experimentally observed transition occurs at a critical
temperature of the order of $J$, signalling the transition itself
to be driven by the intra-layer exchange interaction; details of such
interaction, such as possible easy-axis or easy-plane anisotropies, are hence
fundamental in the analysis of the critical behaviour.  

Several works (see Ref.\onlinecite{Kanamori63} for a review) have shown that,
besides the superexchange interaction, there exist further interaction
mechanisms whose effects may be taken into account by inserting proper
anisotropy terms in the magnetic Hamiltonian; in particular,
the transition observed in  K$_2$NiF$_4$\cite{BirgeneauGS70} ($S=1$),
Rb$_2$FeF$_4$\cite{BirgeneauGS70} ($S=2$), 
K$_2$MnF$_4$\cite{BirgeneauGS73}, Rb$_2$MnF$_4$\cite{BirgeneauGS70} 
($S=5/2$), and others, is 
seen to be possibly due to an easy-axis anisotropy. Such anisotropy  
has often been described in the literature via an external
staggered magnetic field: This choice, despite allowing a qualitative
description of the experimental data, lacks the fundamental property
of describing a genuine phase transition,
as the field explicitly breaks the symmetry and makes the model ordered at 
all temperatures.

In order to produce a second order transition known to be due to an 
easy-axis anisotropy, it is actually appropriate to insert such
anisotropy in the form of an exchange one, thus preserving the symmetry under
inversion along the easy-axis, whose spontaneous breaking manifests itself in
the transition: subject of this paper is the study of the thermodynamic
properties of the resulting model, hereafter called  
Easy-Axis Quantum Heisenberg Antiferromagnet (EA-QHAF).

In Sec.\ref{s.Model} we define the model and discuss its 
general properties, posing the problems we want to address. The 
method used is briefly described in Sec.\ref{s.Results}, where 
results for several thermodynamic quantities and different values of 
the anisotropy are shown and commented. In Sec.\ref{s.Comparison} 
we compare our results with the available experimental data for the
staggered magnetization and susceptibility, and for the correlation length,
of the $S=5/2$ real compound Rb$_2$MnF$_4$\cite{LeeEtal98,BirgeneauGS70}, 
while conclusions are drawn in Sec.\ref{s.Conclusions}.

\section{Model and Method}
\label{s.Model}

The EA-QHAF on the square lattice is  described by the Hamiltonian
\begin{equation}
\hat{\cal H}={J\over 2}\sum_{{\bf i},{\bf d}}\left[\mu\left(
\hat{S}_{\bf i}^x\hat{S}_{{\bf i}+{\bf d}}^x+
\hat{S}_{\bf i}^y\hat{S}_{{\bf i}+{\bf d}}^y\right)+
\hat{S}_{\bf i}^z\hat{S}_{{\bf i}+{\bf d}}^z\right]
\label{e.EAQHAF}
\end{equation}
where ${\bf i}=(i_1,i_2)$ runs over the sites of a square lattice,
${\bf d}$ connects each site to its four nearest neighbours,
$J>0$ is the antiferromagnetic exchange integral and $\mu$ is the anisotropy
parameter ($0\le\mu<1$ for easy-axis models). The spin operators 
$\hat{S}^\alpha_{\bf i}$ ($\alpha=x,y,z$) are such that $|\hat{\bf
S}|^2=S(S+1)$ and obey 
$[\hat{S}^\alpha_{\bf i},\hat{S}^\beta_{\bf j}]=\
i\varepsilon_{\alpha\beta\gamma}\delta_{{\bf i}{\bf j}}\hat{S}^\gamma_{\bf i}$.

When $\mu=1$ the model looses its easy-axis character and reduces to the
isotropic QHAF. 

The $\mu=0$ case will be hereafter called Ising {\it limit}, not to
be confused with the genuine Ising model, reproduced by Eq.(\ref{e.EAQHAF})
with $\mu=0$ and $S=1/2$. Despite being a very particular case of 
Eq.(\ref{e.EAQHAF}), the two dimensional Ising model on the square lattice
is a fundamental point of reference for the study of the thermodynamic 
properties of the EA-QHAF.


In particular, a renormalization-group analysis of the classical 
counterpart of the model Eq.(\ref{e.EAQHAF})\cite{BanderM88,Roscilde99}  
foresees the occurrence of an Ising-like transition at 
finite temperature for any value of $\mu$, no matter how near to the 
isotropic value $\mu=1$;
this analysis is supported by several works based on classical Monte 
Carlo simulations\cite{PattersonJ71,BinderL76,SerenaGL93,GouveaEtal99}. 

In the quantum case, however, 
no information is given about the value of the critical temperature
$T_{\rm c}(\mu,S)$ as a function of the anisotropy and of the spin, 
save the fact that $T_{\rm c}(0,1/2)=0.567$ \hskip 10pt
 \cite{Onsager44} and $T_{\rm c}(1,S)=0$. We do not hence know
whether or not the small anisotropy ($|\mu-1|\simeq 10^{-2}$) observed in real
compounds can be responsible of transitions occurring at critical temperatures
of the order of $J$, given also the fact that we expect quantum fluctuations
to lower the critical temperature with respect to the classical case. 

We have developed a quantitative analysis of several thermodynamic
properties of the model, by means of the 
semiclassical method called pure-quantum self-consistent harmonic 
approximation (PQSCHA)\cite{CTVV92,CGTVV95,CTGMV99}, already
succesfully applied to many magnetic systems in one and 
two dimensions. The method reduces quantum expressions for statistical  
averages to effective classical-like ones, containing temperature and 
spin-dependent
renormalization parameters. The thermodynamics of the effective  
model can then be studied by means of classical techniques,
 like Monte Carlo simulations.

The PQSCHA is known to be particularly suitable for anisotropic 
(easy-plane or easy-axis) models,
 but it also gives very good results in the isotropic QHAF
so that we can confidently use it to investigate the possible 
crossover between
a Heisenberg-like behaviour at high temperature and an Ising-like one
near $T_{\rm c}$; such crossover is detected in the
experimental data to such an extent that, well above the transition, 
real compounds can be satisfactorily described by the isotropic QHAF .

\section{Results}
\label{s.Results}

The main output of the PQSCHA is the effective Hamiltonian 
appearing in all statistical averages, whose expression for the
model described by Eq.(\ref{e.EAQHAF}) reads 
 \begin{eqnarray}
{\cal H}_{\rm eff}&=&
-{1\over 2}J\widetilde{S}^2\sum_{{\bf i},{\bf d}}\left[
\theta^4_\perp\mu
\left(s_{\bf i}^xs_{{\bf i}+{\bf d}}^x+s_{\bf i}^ys_{{\bf i}+{\bf d}}^y\right)+
\theta^2_\parallel\theta^2_\perp 
s_{\bf i}^zs_{{\bf i}+{\bf d}}^z\right]\nonumber \\
&+&{\cal{G}}(t,\widetilde{S})~,
\label{e.Heff}
\end{eqnarray}
where ${\bf s}=(s^x,s^y,s^z)$ is a classical unit vector, 
$\widetilde{S}=S+1/2$, and $t=T/J\widetilde{S}^2$ is the reduced temperature 
hereafter used. The appearance of the minus sign in front of the effective
Hamiltonian is due to the fact that in Eq.(\ref{e.Heff}), as in all the
classical-like expressions reported below, spins belonging to one of the two
sublattices have been flipped, being this an innocuous operation at a classical
level.

The renormalization parameters are
\begin{equation}
\theta^2_\parallel=1-{{\cal D}_\parallel\over 2}~~~~~,~~~~~
\theta^2_{\perp}=1-{{\cal D}_\perp\over 2}~,
\label{e.theta}
\end{equation}
where the coefficients 
\begin{eqnarray}
{\cal D}_\parallel&=&{1\over N\widetilde{S}}\sum_{\bf k}
{a_{\bf k}\over b_{\bf k}}\left(1-\mu\gamma_{\bf k}\right){\cal L}_{\bf k} 
~,\nonumber\\
{\cal D}_\perp&=&{1\over N\widetilde{S}}\sum_{\bf k}
{a_{\bf k}\over b_{\bf k}}\left(1-{\gamma_{\bf k}\over\mu}\right)
{\cal L}_{\bf k} 
\label{e.D}
\end{eqnarray}
are self-consistently determined by solving Eqs.(\ref{e.theta}) and
(\ref{e.D}), with
\begin{equation}
a^2_{\bf k}=4\left(\theta^2_\parallel+\mu\theta^2_\perp\gamma_{\bf k}\right),~
b^2_{\bf k}=4\left(\theta^2_\parallel-\mu\theta^2_\perp\gamma_{\bf k}\right),
\label{e.akbk}
\end{equation}
and 
$$
\gamma_{\bf k}={1\over 4}\sum_{\bf d}\esp{i{\bf k}\cdot{\bf d}},~~
{\cal L}_{\bf k}=\coth f_{\bf k}-{1\over f_{\bf k}},~~
f_{\bf k}={a_{\bf k}b_{\bf k}\over 2\widetilde{S}t}~,
$$
{\bf k} being the wave vector in the first Brillouin zone.

The temperature and spin dependent
uniform term 
$$
{\cal{G}}(t,\widetilde{S})=
2J\widetilde{S}^2(1-\theta^2_\parallel\theta^2_\perp) 
 + J\widetilde{S}^2t
\sum_{\bf k}\ln\left({\sinh f_{\bf k}\over \theta^2_\perp f_{\bf k}}\right)
~, 
$$
does not enter the expressions for the statistical averages, but
contributes to the free energy and to the related thermodynamic quantities.

The renormalization coefficients 
${\cal{D}}_\parallel$ and ${\cal{D}}_\perp$        
measure the pure-quantum fluctuations  
parallel and perpendicular to the 
easy-axis, respectively, of one spin with respect to its nearest neighbours,
and vanish in both the high-temperature
and the classical $S\to\infty$ limit.
It is worthwhile noticing that the fluctuations along the easy-axis only 
enters the renormalization of the $z$-component of the exchange interaction.

Defining the effective exchange integral and effective anisotropy 
$$
J_{\rm eff}=J~\theta^2_\parallel\theta^2_\perp ~~~,~~~
\mu_{\rm eff}=\mu~{\theta^2_\perp\over\theta^2_\parallel}
$$
Eq.(\ref{e.Heff}) can be written in the form
\begin{eqnarray}
{\cal H}_{\rm eff}&=&
-{1\over 2}J_{\rm eff}\widetilde{S}^2\sum_{{\bf i},{\bf d}}\left[
\mu_{\rm eff}\left(s_{\bf i}^xs_{{\bf i}+{\bf d}}^x+
s_{\bf i}^ys_{{\bf i}+{\bf d}}^y\right)+
s_{\bf i}^zs_{{\bf i}+{\bf d}}^z\right]\nonumber \\
&+&{\cal G}(t,\widetilde{S})~,
\label{e.Heff1}
\end{eqnarray}
to make evident that the PQSCHA leads 
to a classical-like effective model of the same form of 
the original quantum one, whose thermodynamic properties can be 
studied by classical numerical technique,
properly taking into account the spin and temperature dependence of the
renormalized parameters $J_{\rm eff}$ and $\mu_{\rm eff}$.
As $\theta^2_\parallel<\theta^2_\perp<1$,
it is $J_{\rm eff}<J$ and $1>\mu_{\rm eff}>\mu$, so that quantum effects
are seen to cause the weakening of both the exchange interaction and the
easy-axis anisotropy; in the isotropic limit
${\cal D}_\parallel={\cal D}_\perp$ and $\mu_{\rm eff}=\mu=1$.

The PQSCHA expression for the statistical average of a quantum operator
$\hat{O}$ is
\begin{equation}
\langle\,\hat{O}\,\rangle={1\over {\cal Z}_{\rm eff}}
\int~d^{\scriptscriptstyle N}\!{\bf s}~O_{\rm eff}~\esp{-\beta{\cal H}_{\rm eff}}
\label{e.aveO}
\end{equation}
where $\beta=T^{-1}$,
${\cal Z}_{\rm eff}=\int d^{^N}\!{\bf s}\,\esp{-\beta{\cal H}_{\rm eff}}$;
$O_{\rm eff}$ is determined by the same procedure leading to 
${\cal{H}}_{\rm eff}$ and contains temperature and spin dependent
renormalizations; after Eq.(\ref{e.aveO}) 
most thermodynamic quantities may be written in
a particularly suggestive form in terms of classical-like statistical
averages defined by the effective Hamiltonian
$$
\langle\,\cdots\,\rangle_{\rm eff}={1\over {\cal Z}_{\rm eff}}
\int~d^{\scriptscriptstyle N}\!{\bf s}~(\,\cdots\,)~\esp{-\beta{\cal H}_{\rm eff}}~;
$$
as far as the evaluation of $\langle\,\cdots\,\rangle_{\rm eff}$ 
is concerned, one must keep in mind that,
because of the temperature dependence of the effective anisotropy appearing
in ${\cal H}_{\rm eff}$, each point in temperature corresponds to a
different effective model. This means that, if the classical Monte Carlo 
technique is used, as done in this work, the 
simulated model changes with temperature, so that, at variance with the 
isotropic case, no existing classical data can be used and a complete series 
of {\it ad hoc} simulations must be carried on. Nevertheless, 
the computational effort required is still that of a classical 
simulation, as the evaluation of the renormalized parameters is a matter of
a few seconds on a standard PC.

The application of the PQSCHA to the EA-QHAF leads to the following 
results for the thermodynamic quantities we have considered:

\begin{itemize}
\item
{\it Internal energy}\\
$u\equiv\langle\hat{\cal H}\rangle/(NJ\widetilde{S}^2)$:
\begin{equation}
u={1\over NJ\widetilde{S}^2}
\langle{\cal H}_{\rm eff}\rangle_{\rm eff}-{\cal G}(t,\widetilde{S})
+{\cal F}(t,\widetilde{S})~,
\label{e.u}
\end{equation}
where 
$$
{\cal F}=-2(\theta^2_\perp-\theta^2_\parallel)
\left[\theta^2_\perp+{2\over 1-\mu^2}(\theta^2_\perp-\theta^2_\parallel)\right]
$$
is a negative zero-point quantum correction term
due to the anisotropy (it vanishes in the isotropic limit).

\item
{\it Staggered magnetization}\\
$m\equiv \sum_{\bf i}(-1)^{i_1+i_2}
\langle\hat{S}^z_{\bf i}\rangle/N\widetilde{S}$:
\begin{equation}
m=\theta^2_\perp\langle s^z_{\bf i}\rangle_{\rm eff}+{\cal M}(t,\widetilde{S})~;
\label{e.m}
\end{equation}
where the magnetization renormalization is seen to be 
due to an effective spin reduction 
($\theta^2_\perp<1$)) and to the appearance of the negative term 
${\cal M}=-(\theta^2_\perp-\theta^2_\parallel)/(1-\mu^2)$ 
which vanishes in the isotropic limit and is finite for $\mu=0$.

\item
{\it Staggered correlation function} above $T_{\rm c}$ 
\noindent
$G(r)\equiv(-1)^{r_1+r_2}\langle\hat{\bf S}_{\bf i}\cdot\hat{\bf S}_{\bf i+r}\rangle
/\widetilde{S}^2$
with ${\bf r}=(r_1,r_2)$ any vector of the square lattice 
and $r=|{\bf r}|$:
\begin{equation}
G(r)=\theta^4_{\bf r}
\langle{\bf s}_{\bf i}\cdot{\bf s}_{{\bf i}+{\bf r}}\rangle_{\rm eff}~,
\label{e.Gr}
\end{equation}
where $\theta^2_{\bf r}=1-{\cal{D}}_{\bf r}/2$ and 
$$
{\cal{D}}_{\bf r}={1\over N\widetilde{S}}\sum_{\bf k}{a_{\bf k}\over b_{\bf k}}
\left(1-\cos({\bf k}{\cdot}{\bf r})\right){\cal L}_{\bf k}
$$
is a further site-dependent renormalization coefficient.
For increasing $r$, the coefficient ${\cal{D}}_{\bf r}$ 
rapidly converges to a uniform term,
so that the asymptotic ($r\to\infty$) behaviour
of $G(r)$ is actually determined by that of the effective classical-like
correlation function 
$\langle{\bf s}_{\bf i}\cdot{\bf s}_{{\bf i}+{\bf r}}\rangle_{\rm eff}$.

\item
{\it Staggered static susceptibility} above $T_{\rm c}$
\noindent $\chi\equiv\sum_{\bf r}G(r)/3$:
\begin{equation}
\chi={1\over 3}
\left[ {S(S+1)\over \widetilde{S}^2}+\sum_{{\bf r}\neq 0}G({r})\right]
\label{e.chi}
\end{equation}

\item
{\it Correlation length} above $T_{\rm c}$
\noindent
We have determined the correlation length $\xi$ by fitting $G(r)$ with the
expression proposed by Serena,Garcia and Levanyuk\cite{SerenaGL93} 
\begin{equation}
G(r)\propto{1\over\xi^{1\over 4}}{\esp{-r/\xi}\over 
\left(r/\xi\right)^{1\over 2}+\left(r/\xi\right)^{1\over 4}}
\label{e.SGL}
\end{equation}
which interpolates the two asymptotic behaviours $r\to\infty$ and $r\to 0$
of the Ising model\cite{KadanoffEtal67}.

\end{itemize}

In what follows, we show our results as obtained combining the above
PQSCHA expressions with the numerical output of the classical Monte Carlo
simulations we have performed to evaluate the effective statistical averages
$\langle\,\cdots\,\rangle_{\rm eff}$. At variance with the isotropic case, 
where results for different values of the spin are obtained by
the same series of classical simulations, 
we now have to fix the value of the spin in order to determine, for a given
value of $\mu$, the corresponding $\mu_{\rm eff}$ to be used in the simulation.
We have hence concentrated ourselves on the case $\mu=0.9942$ and $S=5/2$,
being these the anisotropy and spin values corresponding to 
the real compound Rb$_2$MnF$_4$;
the more anisotropic case $\mu=0.7$ and $S=5/2$ 
has also been considered, because of its expectedly 
more marked Ising-like features.
\begin{figure}[hbt]
\centerline{\psfig{bbllx=4mm,bblly=10mm,bburx=195mm,bbury=160mm,%
figure=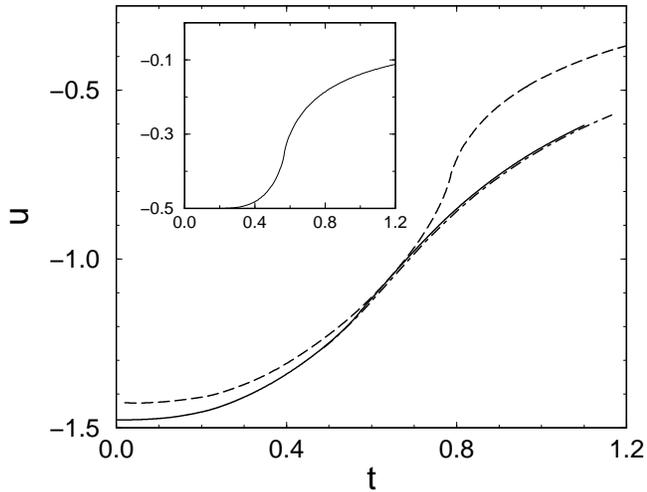,width=85mm,angle=0}}
\caption{Internal energy {\it vs} $t$ for $S=5/2$ and $\mu=0.7$ (dashed), 
$0.9942$ (full), $1$ (dash-dotted); the curve relative to the Ising model
($S=1/2$ and $\mu=0$) is shown in the inset.} 
\label{f.u}
\end{figure}

\begin{figure}[hbt]
\centerline{\psfig{bbllx=4mm,bblly=10mm,bburx=195mm,bbury=160mm,%
figure=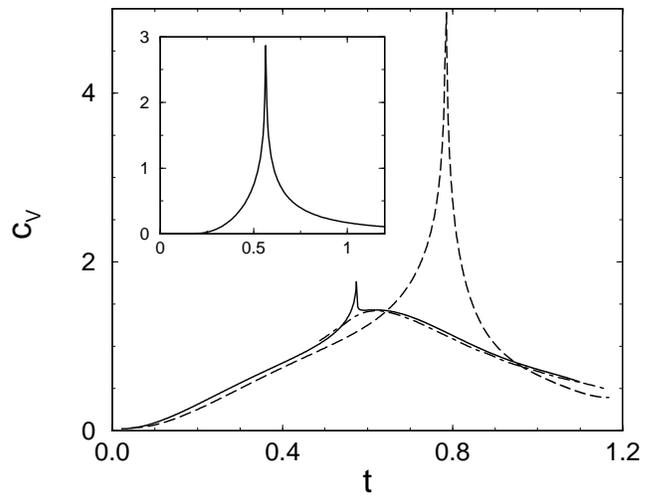,width=85mm,angle=0}}
\caption{Specific heat {\it vs} $t$
(lines and inset as in Fig.\protect{\ref{f.u}}).}
\label{f.sh}
\end{figure}

In Fig.\ref{f.u} and \ref{f.sh} we show the internal energy and specific heat,
versus temperature, for both
values of $\mu$, compared with the isotropic case\cite{CTVV9697iso} $\mu=1$.
As the value of the critical temperature increases
for larger anisotropy, but diminishes when smaller spin values are considered,
we avoid the confusing direct comparison between our curves ($\mu>0$,$S=5/2$)  
and the correponding quantities for the Ising model ($\mu=0$,$S=1/2$),
by showing the latter as Insets.   

The Ising character of the $\mu=0.7$ model is evidenced by 
the pronounced peak in the specific heat, corresponding to a 
qualitatively different temperature dependence of the internal energy  
with respect to that of the isotropic model. 
Although such difference is almost not perceptible when the 
$\mu=0.9942$ curve is considered, a clear peak in the specific heat is
still present, testifying to a persistence of the Ising-like behaviour
even in this {\it quasi}-isotropic model.

\begin{figure}[hbt]
\centerline{\psfig{bbllx=4mm,bblly=10mm,bburx=195mm,bbury=160mm,%
figure=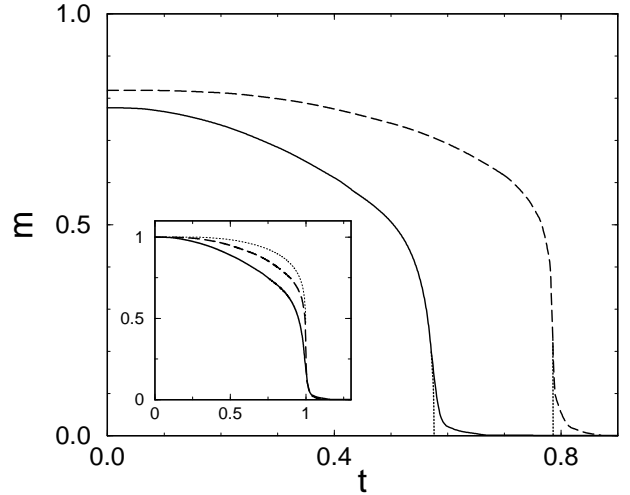,width=85mm,angle=0}}
\caption{Staggered magnetization {\it vs} $t$ for $S=5/2$ and $\mu=0.7$
(dashed), $0.9942$ (full); the dotted lines are the data fits 
used to extract the critical temperature values.
In the inset the same curves,
normalized to the saturation value,
are shown versus $t/t_{\rm c}$, together with the magnetization of the Ising
model (dotted).}
\label{f.m}
\end{figure}

The same conclusion is drawn when the 
staggered magnetization is considered: in Fig.\ref{f.m}
we see that for $\mu=0.9942$ there exist a wide temperature range where
the system is ordered, with a critical temperature $t_{\rm c}$ that, despite
being lower than the one relative to the $\mu=0.7$ case, is still of the order
of $J$. In the inset we show the magnetization curves
normalized to their saturation values, as functions of $t/t_{\rm c}$,
together with that of the Ising model:
It is evident that an increased anisotropy
causes a sharpening of the way the magnetization vanishes.

The critical temperatures hereafter used are $t_{\rm c}=0.785$ 
for $\mu=0.7$ and $t_{\rm c}=0.575$ for 
$\mu=0.9942$; these values have been determined 
by locating at best the correlation length divergence and 
consistently coincide with those emerging from the analysis of the critical
behaviour of other quantities. For instance the critical temperatures
determined by fitting the magnetization curves (see the dotted 
lines in Fig.\ref{f.m}) are $t_{\rm c}'=0.787$ for
$\mu=0.7$ and $t_{\rm c}'=0.576$ for $\mu=0.9942$. 

It is to be noticed that even below $t_{\rm c}$, where the finite value
of the magnetization would suggest a complete predominance of the Ising
character, the system does actually display features which are distinctive 
of the isotropic model. 
In particular, the specific heat for both values of $\mu$
shows, after an exponential start typical of a gapped dispertion relation,
a change in the curvature and an almost linear temperature dependence,
due to the excitation of long-wavelength low-energy modes of the same type 
of those characterizing the isotropic model.
It is just in the vicinity of $t_{\rm c}$ that a new change in the curvature
announces the forthcoming transition.

\begin{figure}[hbt]
\centerline{\psfig{bbllx=4mm,bblly=10mm,bburx=195mm,bbury=160mm,%
figure=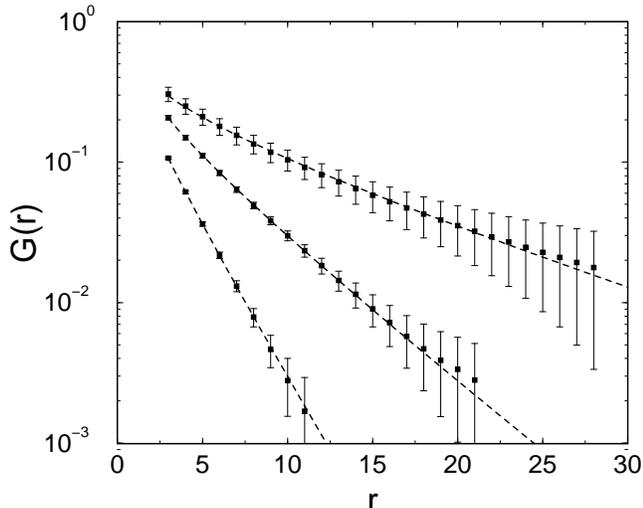,width=85mm,angle=0}}
\caption{Correlation functions for $S=5/2$, $\mu=0.9942$  
and $t=0.6$, $0.7$, $0.85$ (from the top)).}
\label{f.Gr}
\end{figure}

In Fig.\ref{f.Gr} we show the correlation function $G(r)$ as a function of $r$
in the quasi-isotropic case $\mu=0.9942$ and for three different temperatures.
The fit of our data with the Serena-Garcia-Levanyuk 
function Eq.(\ref{e.SGL}) is very good in the whole temperature range
examined,   
and hence leads to a clean evaluation of the correlation length.

\begin{figure}[hbt]
\centerline{\psfig{bbllx=4mm,bblly=10mm,bburx=195mm,bbury=160mm,%
figure=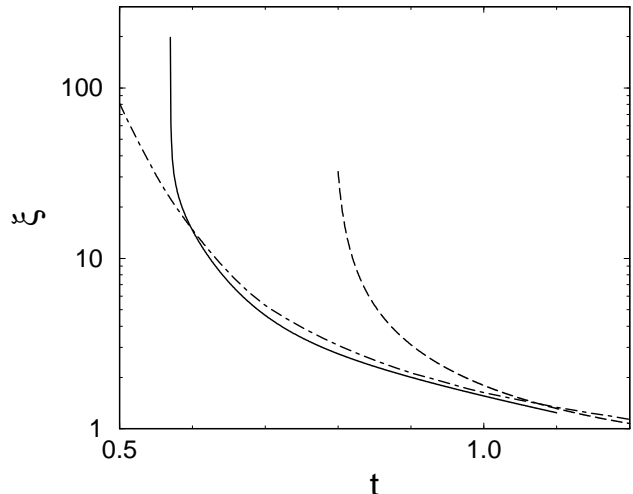,width=85mm,angle=0}}
\caption{Correlation length {\it vs} $t$ for $S=5/2$ and $\mu=0.7$ (dashed),
$0.9942$ (full), $1$ (dash-dotted).}
\label{f.xi}
\end{figure}


In Fig.\ref{f.xi} we
show the correlation length and also report the curve for the 
isotropic model\cite{CTVV9697iso}:
we notice that the $\mu=0.9942$ curve lays on the isotropic one up to
correlation lengths of the order of $20$ lattice spacings 
(i.e. $t\simeq 1.03 t_{\rm c}$), while for $\mu=0.7$ a
deviation is evident already for $\xi\simeq 2$ 
(i.e. $t\simeq 1.3 t_{\rm c}$).
This means that, in the former case, there is a temperature region
where the model significantly behaves like the isotropic model,
as far as the correlation length is concerned,  
and it is hence meaningful to introduce the idea of a crossover from
a Heisenberg- to an Ising-like regime\cite{LeeEtal98}; on the contrary, 
when $\mu=0.7$ the Ising character of the model is manifest already 
when correlations over a few lattice spacings develop.

\section{Comparison with experimental data}
\label{s.Comparison}

The results shown in previous section 
qualitatively explain the mechanism possibly underlying the 
finite-temperature ordering experimentally
observed in many {\it quasi}-two dimensional real magnets.

As for a more precise quantitative analysis, 
we have concentrated ourselves on the $S=5/2$
magnet Rb$_2$MnF$_4$:
reason for this choice is the availability of recent neutron
scattering data\cite{LeeEtal98} relative to such compound and the 
fact that,
because of its cristallographic structure, Rb$_2$MnF$_4$ is known to 
behave as a two-dimensional magnet both above and below the observed 
transition\cite{BirgeneauGS70,CowleyEtal77}.
This means that the critical behaviour is not contaminated by the onset
of three-dimensional order and a clean characterization of the 
transition is possible, as well as a meaningful comparison with the 
experimental data for the magnetization below $T_{\rm c}$.

The model parameters $J_{\rm s}=7.62\pm0.09~{\rm K}$ and 
$\mu_{\rm s}=0.9953$ available
in the literature\cite{deWijnWW73} for the compound Rb$_2$MnF$_4$
are obtained by fitting the extrapolated $T\to 0$ experimental data 
for the spin-wave frequencies 
with the expression
\begin{equation}
\omega_{\bf k}=4J_{\rm s}S\sqrt{{1\over\mu_{\rm s}^2}-\gamma_{\bf k}^2}~;
\label{e.omex}
\end{equation}
this means that $J_{\rm s}$ and $\mu_{\rm s}$ are renormalized by 
the zero-point quantum fluctuations and are not the bare
values to be inserted in Eq.(\ref{e.EAQHAF}). These have hence 
been determined equating Eq.(\ref{e.omex}) with the zero-temperature 
dispersion relation relative to the EA-QHAF as given by the PQSCHA
\begin{equation}
\omega_{\bf k}=4J\widetilde{S}\mu\theta_\perp^2(0)
\sqrt{{\theta^4_\parallel(0)\over\mu^2\theta^4_\perp(0)}-\gamma^2_{\bf k}}
\label{e.omSCHA}
\end{equation}
where $\theta^2_\parallel(0)$ and $\theta^2_\perp(0)$ are the renormalization
parameters defined in Eq.(\ref{e.theta}) evaluated at $t=0$.

The resulting equation for $\mu$ 
$$
\mu=\mu_s{\theta^2_\parallel(0)\over \theta^2_\perp(0)}
$$
must be self-consistently solved, as both $\theta_\perp$
and $\theta_\parallel$ depend on $\mu$, and gives
$\mu=0.9942$.

Once $\mu$ is determined, the equation for the exchange integral
$$
J={S\over\widetilde{S}}{J_s\over \mu\theta^2_\perp(0)}
$$
is straightforwardly solved and gives $J=7.42~{\rm K}$;

\begin{figure}[hbt]
\centerline{\psfig{bbllx=4mm,bblly=10mm,bburx=195mm,bbury=160mm,%
figure=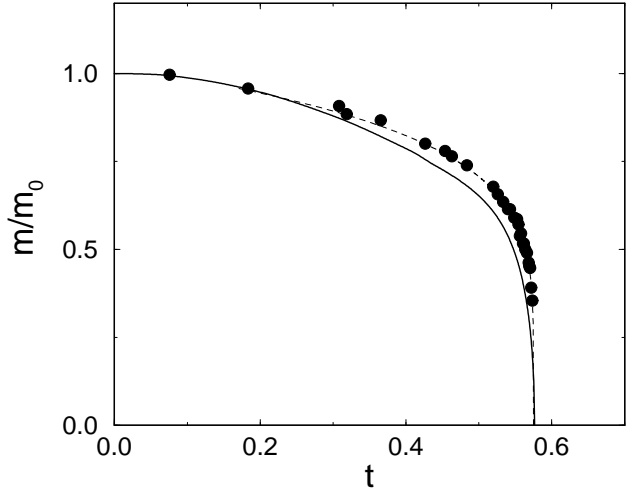,width=85mm,angle=0}}
\caption{Staggered magnetization {\it vs} $t$ for Rb$_2$MnF$_4$,
normalized to the saturation value $m_0$:
our results (full line) are compared with experimental data (full circles) 
from Ref.\protect\onlinecite{BirgeneauGS70}; 
the dashed line is the interpolating curve therein proposed.}
\label{f.mexp}
\end{figure}

In Fig.\ref{f.mexp} our results for the staggered magnetization
are shown together with the experimental data from Ref.\onlinecite{BirgeneauGS70}
and the interpolating curve there proposed.
Besides the overall agreement in the whole ordered phase,
it is to be noticed that our prediction for the value of the critical 
temperature perfectly coincides with the one deriving from the experimental 
analysis, which gives $T_{\rm c}=38.4~{\rm K}$ (i.e. $t_{\rm c}=0.575$);
such a large value of $T_{\rm c}$ with respect to the exchange integral
$J=7.42~{\rm K}$, should not surprise, as the squared
of the spin value has been actually extracted from the latter.

\begin{figure}[hbt]
\centerline{\psfig{bbllx=4mm,bblly=10mm,bburx=195mm,bbury=160mm,%
figure=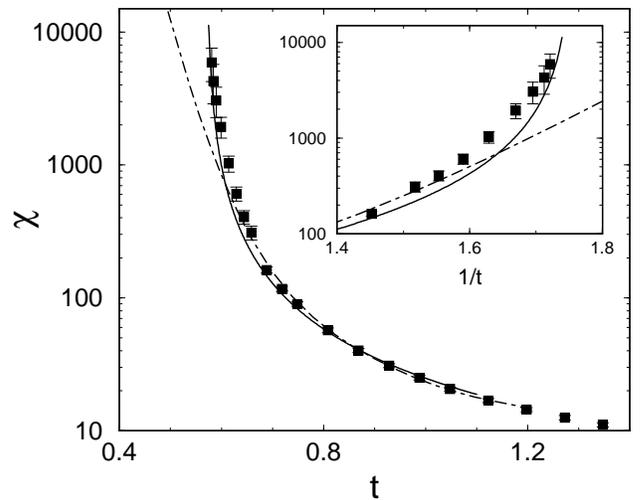,width=85mm,angle=0}}
\caption{Staggered susceptibility {\it vs} $t$ for $S=5/2$, $\mu=0.9942$
(full) and $\mu=1$ (dashed-dotted); symbols are neutron scattering data 
on Rb$_2$MnF$_4$ from Ref.\protect\onlinecite{LeeEtal98}. A zoom of the critical
region is shown in the inset.}
\label{f.chi}
\end{figure}

\begin{figure}[hbt]
\centerline{\psfig{bbllx=4mm,bblly=10mm,bburx=195mm,bbury=160mm,%
figure=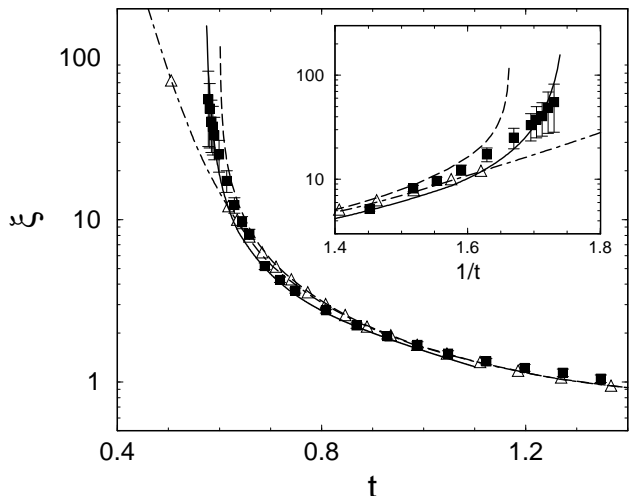,width=85mm,angle=0}}
\caption{Correlation length {\it vs} $t$;
lines, symbols and inset as in Fig.\protect\ref{f.chi} apart from
the dashed curve, representing the result of a mean-field approach 
to the anisotropy proposed in Ref.\protect\onlinecite{LeeEtal98}, and the
triangles, which are Quantum Monte Carlo data for the isotropic model, from
Ref.\protect\onlinecite{BeardCK00}.}
\label{f.xiexp}
\end{figure}

In order to obtain the best quantitative description of the EA-QHAF in the
paramagnetic phase, and given the small anisotropy of Rb$_2$MnF$_4$,
the lowest-temperature data for the staggered susceptibility and 
correlation length, 
shown in Figs.\ref{f.chi} and \ref{f.xiexp}, are 
determined by the PQSCHA version introduced in Ref.\onlinecite{CTVV9697iso} 
and there shown to be the most appropriate to study the isotropic model.
The difference between such version and the original one described in
Sec.\ref{s.Model} consists in the appearance of the renormalization
parameters $\kappa^2_{\parallel,\perp}$, instead of
$\theta^2_{\parallel,\perp}$,
in Eqs.(\ref{e.akbk}), with 
$\kappa^2_{\parallel,\perp}=\theta^2_{\parallel,\perp}-{\cal
D}_{\parallel,\perp}^{\rm cl}/2$,                      
${\cal D}_{\parallel,\perp}^{\rm cl}$ being the renormalization coefficients
determined by the classical self-consistent harmonic approximation.

The agreement between our results and the experimental data is indeed
noticeable, given also the fact that no best-fit procedure has been used.
In addition, our results represent a clear improvement with respect to those 
coming from a mean-field treatment of the anisotropy, proposed by 
Keimer {\it et al.}\cite{KeimerEtal92}, which enables one to derive 
the correlation length of the anisotropic model directly from the data 
of the isotropic one, as done in Ref.\onlinecite{LeeEtal98} starting 
from the PQSCHA results for the isotropic model\cite{CTVV9697iso} itself. 
In particular, it is
evident from Fig.\ref{f.xiexp} that the mean-field approach, apart from
accounting for the existence of the phase transition, leads to an 
overestimate of the critical temperature, while the model with exchange
anisotropy gives a very accurate estimate of $t_c$.

\section{Conclusions}
\label{s.Conclusions}

In this work we have studied the easy-axis quantum Heisenberg antiferromagnet
on the square lattice by means of the pure-quantum self-consistent harmonic
approximation; expressions for several quantum statistical averages 
have been determined for the general model, with any value of the spin and 
of the anisotropy. The numerical work, consisting in classical Monte Carlo
simulations on a properly renormalized model, 
 have been concentrated on the $S=5/2$, $\mu=0.7$ and $S=5/2$, $\mu=0.9942$
cases, the latter corresponding to the real compound Rb$_2$MnF$_4$.

We have shown that a finite temperature transition is present in both cases and
that such transition is cleary of an Ising type; the value of the 
corresponding critical temperature have been determined by the analysis of 
the temperature dependence of the correlation length and has been always found
to perfectly agree with that extracted by the analysis of 
other thermodynamic quantities.

Despite the essential presence of the transition, the EA-QHAF does also
display, both below and above $t_{\rm c}$,
features which are typical of the isotropic model. 
In the ordered phase it is the specific heat behaviour that testify to the
existence of long-wavelength low-energy isotropic-like excitations.
On the other hand,
when the critical region is abandoned in the paramagnetic phase, 
the anisotropy looses its fundamental role and
a crossover towards the isotropic behaviour is observed, at least as far as
the correlation length is concerned; such crossover, however, has a weaker
meaning for larger anisotropy, being confined, already for $\mu=0.7$,
to the high temperature region where $\xi$ is the order of 
the lattice spacing and differences between different models become
irrelevant.

We have compared our theoretical results with the neutron scattering
experimental data for the staggered magnetization, staggered susceptibility and
correlation length of the real compound Rb$_2$MnF$_4$ and found an excellent
agreement both for the overall temperature behaviour and for the value of the
critical temperature. 

We can hence conclude that the experimentally observed
finite temperature transition in Rb$_2$MnF$_4$ is due to an easy-axis
anisotropy in the intra-layer exchange interaction and that, despite the
small value of the anisotropy, the compound shows an Ising-like critical
behaviour.

\section{Acknowledgments}
We acknowledge Prof. R.~J. Birgeneau and  Dr. Y. S. Lee for sending us
preprints and experimental data. 

This work has been partially supported by
the COFIN98-MURST fund.


\end{document}